\documentclass{article}
\usepackage[utf8]{inputenc}
\usepackage{moreverb}

\newcommand{\mytitle}{A task-based approach to parallel parametric linear programming solving, and application to polyhedral computations}

\usepackage{amsthm}

\usepackage{amsfonts,amsmath,algorithm,algpseudocode,multicol}
\usepackage[colorinlistoftodos]{todonotes}
\newcommand{\RR}{\mathbb{R}}
\newcommand{\QQ}{\mathbb{Q}}
\newcommand{\ZZ}{\mathbb{Z}}

\newcommand{\software}[1]{\textsf{#1}}
\newcommand{\none}{\textsf{none}}
\newcommand{\true}{\textsf{true}}
\newcommand{\false}{\textsf{false}}
\newcommand{\ignore}[1]{{}}

\newcommand{\eg}{e.g.}
\graphicspath{{figures/}{xp/}}
\usepackage{paralist,afterpage}

\usepackage[backend=bibtex]{biblatex}
\usepackage{dmbiblatex}
\bibliography{PPLP_ICCS_2019}

\usepackage{subfig}

\theoremstyle{remark}
\newtheorem{example}{Example}

\newenvironment{contexample}{\addtocounter{example}{-1}\begin{example}[continued]}{\end{example}}

\usepackage{authblk}
\usepackage{hyperref}

\title{\mytitle}

\author[1]{Camille Coti}
\author[2]{David Monniaux}
\author[2]{Hang Yu}

\affil[1]{LIPN, CNRS UMR 7030, Universit\'e Sorbonne Paris Nord, 99, avenue Jean-Baptiste Cl\'ement, F-93430 Villetaneuse, France}
\affil[2]{VERIMAG, Univ. Grenoble Alpes, CNRS, Grenoble INP\thanks{Institute of Engineering Univ. Grenoble Alpes}, F-38000 Grenoble, France}

\begin{document}

\maketitle

\begin{abstract}
Parametric linear programming is a central operation for polyhedral computations, as well as in certain control applications.
Here we propose a task-based scheme for parallelizing it, with quasi-linear speedup over large problems.
This type of parallel applications is challenging, because several tasks might
be computing the same region. In this paper, we are presenting the
algorithm itself with a parallel redundancy elimination algorithm, and
conducting a thorough performance analysis. 

\end{abstract}

\section{Introduction}
\label{sec:intro}
A \emph{convex polyhedron} in dimension $n$ is the solution set over $\QQ^n$ (or, equivalently, %
\footnote{Whether emptiness tests, inclusion tests, projection etc. are specified with real or rational variables, the results are the same. It is impossible to distinguish the reals and the rationals using first-order formulas in linear arithmetic.}%
$\RR^n$) of a system of inequalities (with integer or rational coefficients).
Since all the polyhedra we consider here are convex, we shall talk of polyhedra for short.
There also exist integer polyhedra, defined over $\ZZ^n$, but they are very different in many respects; we shall not consider them here.%
\footnote{Among differences, it is possible to check in polynomial time if a given list of inequalities defines an empty polyhedron over $\QQ$ but the same problem is NP-complete over~$\ZZ$.}

Computations over polyhedra arise in low dimension ($n=2$ and $n=3$) for modeling physical objects, but here we are interested in higher dimensions.
Polyhedra in higher dimension are typically used to enclose the set of reachable states of systems whose state can be expressed, at least partially, as a vector of reals or rationals.
For instance, one can study the flow of ordinary differential equations, or more generally the trajectories of hybrid systems, by enclosing the trajectories into a succession of convex polyhedra. % TODO ref
If the internal state of a control system is expressed by a vector of $n$ numeric variables, one may prove that this system never encounters a bad condition by exhibiting a polyhedron $P$ such that the initial state belongs to $P$, no bad condition belongs to $P$, and all possible time steps of the control system leave $P$ stable (that is, it is not possible to execute a step starting in $P$ and ending outside of~$P$).

One generally proves the correctness of programs by exhibiting \emph{inductive invariants} --- an inductive invariant for a loop is a set containing the precondition of the loop, stable by moving  to the next loop iteration, and implying the desired postcondition.
Since providing inductive invariants by hand is tedious, it is desirable to automate that process. One approach for doing so is \emph{abstract interpretation}, where an ascending sequence of sets of states is computed until reaching a fixed point.
One may for instance search such sets as products of intervals, an approach known as \emph{interval analysis}, but the lack of relationships between the dimensions tends to severely limit the kind of properties that can be proved (e.g., one cannot have an invariant $i < n$ where $n$ is a parameter: this is neither an interval on $i$ nor on $n$ nor a combination thereof). 
Cousot and Halbwachs proposed searching for polyhedral inductive invariants \cite{DBLP:conf/popl/CousotH78,Halbwachs_PhD}.
The operations needed there are projection, more generally image by an affine transformation, convex hull, and inclusion (or equality) test, together with an extrapolation operator known as \emph{widening}.

Such usages of convex polyhedra suffered from the \emph{curse of dimensionality}: complexity grows quickly with the number of dimensions---the number of numeric variables in the program or hybrid system under analysis.
This problem is exacerbated in approaches that double the number of variables to represent pairs of states, or add extra variables for representing nonlinear terms (e.g., a variable $v_{xy}$ is added to represent $xy$ \cite{DBLP:conf/vmcai/MarechalFKMP16}).
This led to either restricting polyhedra to subclasses with lower computational costs (e.g., restricting inequalities to $\pm x_i \pm x_j \leq C$ as in the \emph{octagons} \cite{DBLP:journals/lisp/Mine06}),
or severely restricting the number of dimensions of the system under consideration (e.g., by focusing on a subset of interest of the program variables, disregarding relationships with variables outside of that subset).

Why this curse of dimensionality? There exist multiple libraries for computing over polyhedra (\software{NewPolka},%
\footnote{Jeannet's NewPolka is now part of Apron \url{http://apron.cri.ensmp.fr/library/} \cite{DBLP:conf/cav/JeannetM09}}
\software{Parma Polyhedra Library},\footnote{\url{https://www.bugseng.com/ppl} \cite{BagnaraHZ08SCP}}
\software{PolyLib},\footnote{\url{https://icps.u-strasbg.fr/PolyLib/}}
\software{CDD}\dots\footnote{\url{https://www.inf.ethz.ch/personal/fukudak/cdd_home/}}).
They all use the \emph{double description} \cite{Motzkin53_double_description} of convex polyhedra: both as \emph{constraints} (inequalities or equalities) and \emph{generators} (vertices, and, for unbounded polyhedra, lines and rays).
Some operations are indeed simpler on one description than the other, and
Chernikova's algorithm \cite{LeVerge94} converts between the two.
Having both descriptions is also handy for removing redundant constraints and generators, which are produced by many of the algorithms.

The double description has many advantages, but one major weakness.
A very common case of invariant is when one knows the interval of variation of each variable: $[l_1,h_1] \times \dots \times [l_n,h_n]$.
Such a polyhedron has $2n$ constraints ($l_i \leq x_i$ and $x_i \leq h_i$ for $1 \leq i \leq n$) and $2^n$ vertices (each $x_i$ is independently chosen in $\{l_i,h_i\}$).
The double description then blows up.
One workaround for the above case is to detect that the polyhedron is exactly a Cartesian product of simpler polyhedra, and to compute as locally as possible in terms of that product \cite{DBLP:journals/fmsd/HalbwachsMG06}, but this fails if the polyhedron is \emph{almost} a Cartesian product;
We thus chose to completely do away with the generator representation and use constraints only.%
\footnote{By polyhedral duality, which exchanges constraints and generators, the worst-case for generators translates into a worst-case for constraints. The crucial point is that the worst-case for generators occurs very naturally in the analysis of programs or hybrid systems.}

The question was how to compute over convex polyhedra described by constraints only.
It is possible to reduce most operations (image, convex hull etc.) to projection.
An algorithm for projecting polyhedra described by inequalities only, Fourier-Motzkin elimination \cite{Encyclopedia_of_Optimization_Fourier_Motzkin}, has long been known. % \cite{Fourier_1824}.
Unfortunately it tends to generate a very large number of redundant constraints, which must be eliminated using an expensive procedure.
%; even if this expensive procedure is replaced by one \cite{DBLP:conf/vmcai/MarechalP17} that handles cheaply some easy cases.

We instead turned to \emph{parametric linear programming}.
Image, projection, convex hull can all be formulated as solutions to linear programs where parameters occur linearly within the objective~\cite{DBLP:phd/hal/Marechal17,Boulme_et_al_SYNASC2018}:
given a system of equalities and inequalities, defining a convex polyhedron $P$ in higher dimension,
and a parametric bilinear objective function $f(x,\lambda)$ where $x$ is the point and $\lambda$ a vector of parameters, give for each $\lambda$ a vertex $x^*$ of $P$ such that $f(x^*,\lambda)$ is minimal.
A solution to such a program is a quasi-partition of the space of parameters $\lambda$ into convex polyhedra, with one optimum associated to each polyhedron.
The issue is how to compute this solution efficiently.
In this article, we describe how we parallelized our algorithm for doing so.

In addition to computations on convex polyhedra, parametric linear programming is also used for control applications \cite{Jones+Automatica2007-Multiparametric-linear-programming-with-applications-to-control}: instead of using a solver,
whose computation times are high and hard to predict,
inside the control loop,
the solution of the linear program is tabulated as a piecewise linear function over the values of the parameters.
Parametric linear programming is also used for affine linear approximations of nonlinear expressions~\cite{DBLP:conf/vmcai/MarechalFKMP16}.

Parallelizing this type of applications seems straightforward at first
sight, since each polyhedron can be computed independently from the
other ones. However, it is actually challenging, because several
computation units (threads or processes) might
be computing the same region at the same time. In this paper, we present
a parallel redundancy elimination
algorithm that improves the performance by eliminating redundant
computations between concurrent threads,
and conduct a thorough experimental study of the
task-based parallel scheme presented shortly in~\cite{CMY19}. 

%%% Local Variables:
%%% ispell-local-dictionary: "en"
%%% TeX-master: "main"
%%% End:

\section{Related works}
\label{sec:related}
Most libraries for computing over convex polyhedra are based on the double description approach: a polyhedron is described both as the convex hull of its vertices (and, in the case of unbounded polyhedra, rays and/or lines), and as the solution set of a system of equalities and inequalities.
They convert from one description to the other using Chernikova's algorithm~\cite{LeVerge94}, which computes a set of generators (vertices, rays, lines) from a set of (in)equalities (and, dually, the converse) by considering each (in)equality in a sequence and computing the intersection of the polyhedron defined by the previous (in)equalities in the sequence and the current one.
To our best knowledge, there is no parallel version of Chernikova's algorithm,
and we fail to see how to parallelize its main loop.
It may be possible to parallelize the inner loops that compute the generators of the intersection of a polyhedron $P$ and an (in)equality $C$ given the generators of $P$.
An alternative to Chernikova is the reverse search vertex enumeration algorithm \cite{Polytopes_DMV29}.

We also opted out of the double description because it is difficult to independently verify that a polyhedron described by generators
includes the polyhedron that should have been computed.%
\footnote{It is co-NP-hard to check that, given the description of a polyhedron $A$ by constraints and a polyhedron $B$ described by generators, $A$ is included in $B$~\cite{DBLP:journals/mp/FreundO85}.
  Furthermore, the \emph{vertex enumeration} problem, that is, checking whether, given a polyhedron $P$ described by a list of constraints and a set of vertices $V$ of that polyhedron, $P$ has another vertex not in $V$, is NP-hard for unbounded polyhedra~\cite{DBLP:journals/dcg/KhachiyanBBEG08}.
  As of 2008, it was unknown if the same problem for bounded polyhedra, or, equivalently, that of generator enumeration ($V$ may contain rays and lines) for unbounded polyhedra, was also NP-complete (it is known to be in NP).
  No progress seems to have been made on this front since then.

  Enumeration can be done in polynomial time for \emph{simple} polyhedra, those for which a vertex corresponds to exactly one basis~\cite{Polytopes_DMV29} (no degeneracy);
  note that degeneracy is also a major source of complication in our algorithms.}

Such verifiability is desirable for certain applications: when one computes a polyhedron meant to include all reachable states of a system, to prove that no undesirable state can be reached, then it would be catastrophic that this polyhedron excludes some reachable state due to a bug in a library.
Our Verified Polyhedron Library (VPL)%
\footnote{\url{https://github.com/VERIMAG-Polyhedra/VPL}}
\cite{Marechal_Monniaux_Perin_SAS2017,DBLP:phd/hal/Marechal17,DBLP:conf/vmcai/MarechalP17,DBLP:conf/vmcai/MarechalFKMP16,DBLP:phd/hal/Fouilhe15,DBLP:conf/vstte/FouilheB14,Fouilhe_et_al_SAS2013}
provides, in addition to core computations, an optional layer, formally proved correct within the Coq proof assistant, that performs this verification.

VPL implements a constraint-only description (equalities and inequalities) for polyhedra, using two generations of algorithms.
The first generation maps all operations, including convex hull, to projection, performed by Fourier-Motzkin elimination \cite{Encyclopedia_of_Optimization_Fourier_Motzkin,Fourier_1824}.
The second generation maps all operations to parametric linear programming, performed as in Algorithm~\ref{algo:PLP} except that no floating-point solver is used, just an exact-precision implementation of the simplex algorithm.
Furthermore, VPL is implemented in OCaml, which does not currently allow running computations in multiple threads, and its data structures were designed for compatibility with Coq.
For all these reasons, VPL has lower performance than the C++ implementation described in this paper, even with one thread.

Parametric linear programming with the parameters in the objective function is a generalization of vertex enumeration, for which there exist parallel implementations based on \emph{reverse search} \cite{DBLP:journals/mpc/AvisJ18} (vertex enumeration is the case where there are as many independent parameters as there are variables, so that the optimization direction can point to any direction).
A difference of our approach with reverse search is that we store the nodes already traversed in a central location, which they do not have to do.
Jones and Maciejowski \cite{Jones_reverse_search_4177566} applied reverse search to parametric linear programming; they however warn that while they have better asymptotic complexity than other approaches, the constant hidden in the big-O notation is huge and they warn that their approach is likely to be interesting only on larger examples.
In contrast, we base ourselves on an approach already used in a sequential library that is competitive with double description approaches even on problems in moderate dimension~\cite{Boulme_et_al_SYNASC2018}.

%%% Local Variables:
%%% ispell-local-dictionary: "en"
%%% TeX-master: "main"
%%% End:

\section{Sequential algorithms}
\label{sec:algo}
We shall leave out here how polyhedral computations such as projection and convex hull can be reduced to parametric linear programming --- this is covered in the literature \cite{Jones+JOTA08-On-polyhedral-projections-and-parametric-programming,Marechal_Monniaux_Perin_SAS2017} --- and focus on solving the parametric linear programs.

\subsection{Non-parametric linear programming}
A linear program with $n$ unknowns is defined by a system of equations $AX = B$, where $A$ is an $m \times n$ matrix; a solution is a vector $X$ such that $X \geq 0$ on all coordinates and $AX = B$.%
\footnote{There exist more general descriptions of linear programming with upper and/or lower bound constraints on each coordinate of $X$;
our approach generalizes to them.}
The program is said to be \emph{feasible} if it has at least one solution, \emph{infeasible} otherwise.
In a non-parametric linear program one considers an objective $C$: one wants the solution that maximizes $C^T X$.
The program is deemed \emph{unbounded} if it is feasible yet it has no such optimal solution.

\begin{example}
\label{ex:polygon}
  Consider the polygon $P$ defined by $x_1 \geq 0$, $x_2 \geq 0$, $3x_1-x_2 \leq 6$, $-x_1+3x_2 \leq 6$.
  Define $x_3 = 6 - 3x_1 + x_2$ and $x_4 = 6 + x_1 - 3x_2$.
  Let $X = (x_1,x_2,x_3,x_4)$, then $P$ is the projection onto the first two coordinates of the solution set of $AX=B \land X \geq 0$ where
  $A = \begin{bmatrix}
      1 & -3 & 0 & -1 \\
      -3 & 1 & -1 & 0
    \end{bmatrix}$
    and $B = \begin{bmatrix} 6 \\ 6 \end{bmatrix}$.
    %, in other words
 %$\begin{bmatrix}
  %    1 & -3 & 0 & -1 \\
  %    -3 & 1 & -1 & 0
  %  \end{bmatrix} X =
  %  \begin{bmatrix} 6 \\ 6 \end{bmatrix}$.
\end{example}

A linear programming solver takes as input $(A,B,C)$ and outputs ``infeasible'', ``unbounded'' or an optimal solution~$X^*$.
Some linear programming solvers take $(A,B,C)$ and output $X^*$ as exact rational numbers and ensure that the answer is correct.
Most, however, operate fully on floating-point numbers and their final answer may be incorrect: they may answer ``infeasible'' whereas the problem is feasible, or propose ``optimal solutions'' that are not solutions, or are not optimal.

In addition to an exact $X^*$ or floating-point $\tilde{X}^*$ output, the solvers also provide the discrete information of \emph{optimal basis}:
a solution is obtained by the simplex algorithm setting $n-m$ coordinates of $X^*$ to $0$ (known as \emph{nonbasic variables}) and solving for the other coordinates (known as \emph{basic variables}) using $AX^* =B$,
and the solver provides the partition into basic and nonbasic variables it used.
If a floating-point solver is used, it is possible to reconstruct an exact rational point $X^*$ using that information and a library for solving linear systems in rational arithmetic.
One then checks whether it is truly a solution by checking $X^* \geq 0$ on the reconstructed coordinates.

The optimal basis contains even more information: it contains a proof of optimality of the solution!
The system computes the objective function $C^T X$ as $\sum_{i \in N} \alpha_i X_i$ where $N$ is the set of indices of the nonbasic variables,
and concludes that the solution obtained by setting these nonbasic variables to $0$ is maximal because all the $\alpha_i$ are nonpositive.

If $X^*$ is not a solution of the problem (the condition $X^* \geq 0$ fails) or is not optimal, then one can fall back to an exact implementation of the simplex algorithm, possibly starting it from the last basis considered by the floating-point implementation.

\begin{contexample}
  Assume the objective is $C = \begin{bmatrix} 1 & 1 & 0 & 0\end{bmatrix}$, that is, $C^T X = x_1 + x_2$.
  From $AX = B$ we deduce
  $x_1 = 3 -\frac{3}{8} x_3 - \frac{1}{8} x_4$ and
  $x_2 = 3 -\frac{1}{8} x_3 - \frac{3}{8} x_4$.
  Thus $x_1 + x_2 = 6 - \frac{1}{2} x_3 - \frac{1}{2} x_4$.

  Assume $x_3$ and $x_4$ are nonbasic variables and thus set to $0$, then
  $X^* = (x_1,x_2,x_3,x_4) = (3, 3, 0, 0)$.
  It is impossible to improve upon this solution: as $X \geq 0$, changing the values of $x_3$ and $x_4$ can only decrease the objective $o = 6 - \frac{1}{2} x_3 - \frac{1}{2} x_4$.
  This expression of $o$ in terms of the nonbasic variables can be obtained by linear algebra once the partition into basic and nonbasic variables is known.
\end{contexample}

If the linear programming solver uses floating-point arithmetic and is not to be trusted, it is still possible to reconstruct, by pure linear arithmetic, the expression of the objective function as a function of the nonbasic variables and check the signs of the coefficients.

While the optimal value $C^T X^*$, if it exists, is unique for a given $(A,B,C)$, there may exist several $X^*$ for it, a situation known as \emph{dual degeneracy}.

\begin{contexample}
  Assume the objective is $C = \begin{bmatrix} -1 & 0 & 0 & 0\end{bmatrix}$, that is, $C^T X = - x_1$.
  $X^*$ can be chosen to be $(0,0,6,6)$, $(0,2,8,0)$ or any point in between.
\end{contexample}

The same $X^*$ may be described by different bases, a situation known as \emph{primal degeneracy}, happening when more than $n-m$ coordinates of $X^*$ are zero, and thus some basic variables could be used as nonbasic and the converse.

\begin{figure}[t]
  \centering
\includegraphics[height=5cm]{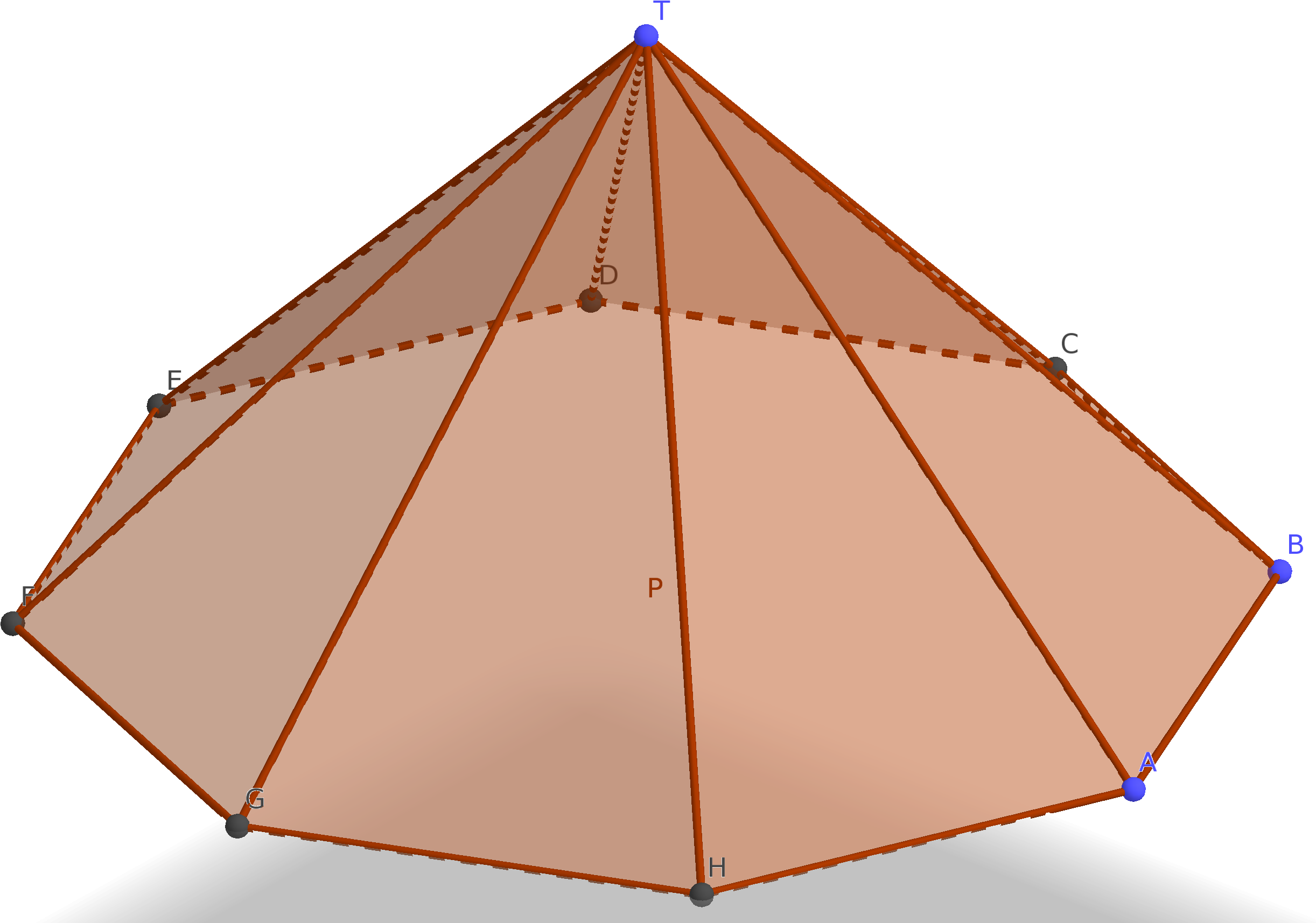}

  \caption{A pyramid based on an octagon, the apex $T$ described by $\binom{8}{3}$ bases.}
  \label{fig:pyramid}
\end{figure}

\begin{example}
Consider the regular polygon $B$ with $k \geq 3$ vertices $\big(1+\cos(2i\pi/k),1+\sin(2i\pi/k)\big) \mid 1 \leq i < k$.
The convex hull of $B$ and $T=(1,1,1)$ is a pyramid with $k+1$ faces
(\eg, $k=4$ is a square pyramid),
defined using $3 + k$ unknowns and $k$ equations
(\eg, for $k=4$, $x_4 = 1 - x_1 - x_2 - x_3$, $x_5 = 1 + x_1 - x_2 - x_3$,
                 $x_6 = 1 - x_1 + x_2 - x_3$, $x_5 = 1 + x_1 + x_2 - x_3$),
plus $x_1,\dots,x_{k+3} \geq 0$.
The apex $T$ of the pyramid corresponds to $(1,1,1,0,\dots,0)$.
It is obtained by picking any 3 variables out of $x_4,\dots,x_{3+k}$ as nonbasic:
there are $\binom{k}{3}$ bases defining it
(Fig.~\ref{fig:pyramid}).
\end{example}

\subsection{Parametric linear programming}
For a \emph{parametric} linear program, we replace the constant vector $C$ by $C_0  + \sum_{i=1}^k \mu_i C_i$ where the $\mu_i$ are parameters.%
\footnote{There exists another, dual, kind of parametric linear programming where the parameters are in the right-hand side $B$. We do not consider it here.}
When the $\mu_i$ change, the optimum $X^*$ changes.
Assume temporarily that there is no degeneracy. Then, for given values of the $\mu_i$, the problem is either unbounded, or there is one single optimal solution~$X^*$.
It can be shown that the region of the $(\mu_1,\dots,\mu_k)$ associated to a given optimum $X^*$ is a convex polyhedron (for $C_0$, a convex polyhedral cone), and that these regions form a quasi partition of the space of parameters (two regions may overlap at their boundary, but not in their interior) \cite{Jones+JOTA08-On-polyhedral-projections-and-parametric-programming,Jones+Automatica2007-Multiparametric-linear-programming-with-applications-to-control,Marechal_Monniaux_Perin_SAS2017}.
The output of the parametric linear programming solver is this quasi-partition, and the associated optima---in our applications, the problem is always bounded in the optimization directions, so we do not deal with the unbounded case.

Let us see in more detail how to compute these regions.
We wish to attach to each basis (at least, each basis that is optimal for at least one vector of parameters) the region of parameters for which it is optimal.

\setcounter{example}{0} % UGLY
\begin{example}[continued]
Instead of $C=\begin{bmatrix} 1 & 1 & 0 & 0\end{bmatrix}$ we consider the parametric $C=\begin{bmatrix} \mu_1 & \mu_2 & 0 & 0 \end{bmatrix}$.
Let us now express $o = C^T X$ as a function of the nonbasic variables $x_3$ and $x_4$:
\begin{equation}
o = (3 \mu_1 + 3 \mu_2) +
    \left(-\frac{3}{8} \mu_1 -\frac{1}{8} \mu_2\right) x_3 +
    \left(-\frac{1}{8} \mu_1 -\frac{3}{8} \mu_2\right) x_4
\end{equation}
The coefficients of $x_3$ and $x_4$ are nonpositive if and only if
$3\mu_1 + \mu_2 \geq 0$ and $\mu_1 + 3\mu_2 \geq 0$, which define
the cone of optimality associated to that basis and to the optimum
$X^* = (3,3,0,0)$.
\end{example}

Note that the description of the cone (or polyhedron) of optimality by the constraints obtained from the sign conditions in the objective function may contain redundant constraints, that is, constraints that can be removed without changing the cone.
It is desirable to remove these.
Some procedures for removing redundant constraints from the description of a region $R_1$ also provide a set of vectors outside of $R_1$: for each constraint in the description they provide a vector violating it~\cite{DBLP:conf/vmcai/MarechalP17},
a feature that will be useful.

Assume now we have solved the optimization problem for a value $C(D)$ of the optimization direction, for a vector of parameters $D_1$, and obtained a region $R_1$ in the parameters (of course, $D_1 \in R_1$).
We now pick $D_2 \notin R_1$  --- if the redundancy elimination procedure provided us with a set of vectors outside of $R_1$, we can store them in a ``working set'' to be processed and choose $D_2$ in it.
We compute the region $R_2$ associated to $D_2$.
Assume that $R_2$ and $R_1$ are adjacent, meaning that they have a common boundary.
We get vectors outside of $R_2$ and add them to the working set.
We pick $D_3$ in the working set, check that it is not covered by $R_1$ or $R_2$, and, if it is not, compute a region $R_3$, etc.
This amounts to a traversal of the adjacency graph of the optimality regions.
The algorithm terminates when the working set becomes empty, meaning the $R_1,\dots$ produced form the desired quasi-partition.

\begin{figure}[t]
	\centering
	\includegraphics[width=5cm]{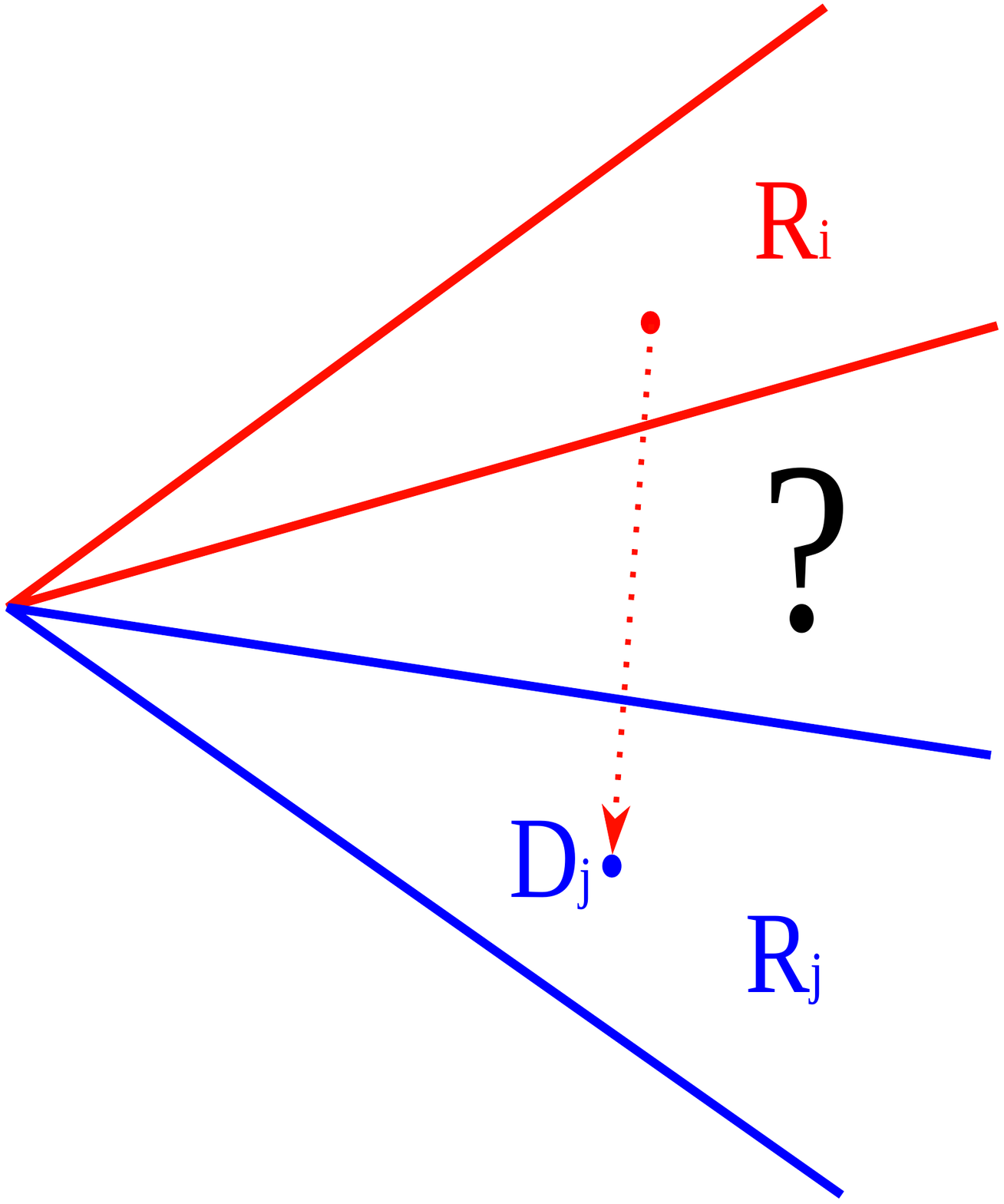}
	\caption{$R_i$ and $R_j$ are not adjacent. The intermediate region is missed.}
	\label{fig:fake_optimum}
\end{figure}

This simplistic algorithm might fail because it assumes that it is discovering the adjacency relation of the graph.
The problem is that, if we move from a region $R_i$ to a vector $D_j \notin R_i$, it is not certain that the region $R_j$ generated from $D_j$ is adjacent to it, so we could miss some intermediate region (Fig.~\ref{fig:fake_optimum}).
In order to cope with this issue, we modify our traversal algorithm as follows.
The working set contains pairs $(R,D')$ where $R$ is a region and $D' \notin R$ a vector (there is also a special value \none\ for $R$).
The region $R'$ corresponding to $D'$ is computed.
If $R$ and $R'$ are not adjacent, then a vector $D''$ in between $R$ and $R'$ is computed, and the pair $(R,D'')$ is added to the working set.
This ensures that at the end, we obtain a quasi-partition.
An additional benefit is that we can obtain a spanning tree of the region graph, with edges from $R$ to $R'$, tracking which region led to which other one.

The last difficulty is degeneracy.
So far we have assumed that each optimization direction corresponded to exactly one optimal vector $X^*$, and that this optimal vector is described by exactly one basis.
This is not the case in general; in this case, the interiors of the optimality regions may overlap.
This situation results in a performance hit; also the final result is no longer a quasi-partition, but instead just a covering of the parameter space (for each possible vector $D$ of parameters, there is at least one region that covers it).
Of course, the value of the objective function $C(D)^T X^*$ must be the same for all optimal vectors $X^*$ associated with the regions covering~$D$.
A covering suffices however for the correctness of our projection, convex hull etc. algorithms.

We are currently investigating approaches for getting rid of degeneracy --- enforcing one optimal vector $X^*$ and only one optimal basis per vector $D$, except at the boundaries.
The methods for doing so rely on lexicographic orderings or perturbations on the objective and/or constant term, or pivoting rules~\cite{Jones+Automatica2007-Multiparametric-linear-programming-with-applications-to-control}.
We have recently proposed a working solution to degeneracy \cite{DBLP:conf/sas/YuM19} but there is still room for improvement.

\begin{algorithm}
  \caption{Sequential parametric linear programming solver.}

$\mathit{float\_lp}(A,B,C)$ is an external procedure returning the optimal basis for maximizing $C^T X$, $AX = B$, $X \geq 0$; it may provide incorrect results.

$\mathit{exact\_lp}$ returns the exact optimum and optimal basis.

$\mathit{midpoint}(R,R',D')$, where $D' \in R'$, computes a vector in between regions $R$ and $R'$.

$\mathit{exact\_point}$ computes the exact rational $X^*$ point corresponding to the basis.

$\mathit{exact\_objective}$ computes the objective function as a bilinear function of the parameters and the nonbasic variables (the output is a matrix).

$\mathit{sign\_conditions}$ translates it into sign conditions on the parameters, defining a cone.

$\mathit{eliminate\_redundancy}(S)$ returns $(R, D_{\mathrm{next}})$, where $R$ is an irredundant set of inequalities defining the same cone as $S$ and $D_{\mathrm{next}}$ are vectors outside of that cone, each violating one different inequality in $R$ but not the others.%
\label{algo:PLP}

% \begin{multicols}{2}\noindent%
\begin{algorithmic}
\Procedure{PLP}{$A,B,C$}
\State pick any nonzero vector of parameters $D_0$
\State $W \gets \{ (\none, D_0) \}$
\State $\mathit{regions} \gets \emptyset$
\While{$W \neq \emptyset$}
\State Pick $(R_{\mathrm{from}},D)$ in $W$ and remove it from $W$
  \State $R_{\mathrm{cov}} \gets \mathit{is\_covered}(D,\mathit{regions})$
%  \If{$R_{\mathrm{cov}} \neq \none$}
%    \State $D' \gets \mathit{midpoint}(R_{\mathrm{from}},R_{\mathrm{cov}},D)$
%    \State $W \gets W \cup \{ (R_{\mathrm{from}}, D') \}$
%  \Else
  \If{$R_{\mathrm{cov}} == \none$}
    \State $\mathit{basis} \gets \mathit{float\_lp}(A,B,C(D))$
    \State $X^* \gets \mathit{exact\_point}(\mathit{basis})$
    \State $o \gets \mathit{exact\_objective}(\mathit{basis})$
    \If{$\neg (X^* \geq 0 \land o \leq 0)$}
      \State $(\mathit{basis},X^*) \gets \mathit{exact\_lp}(A,B,C(D))$
    \EndIf
    \State $S \gets \mathit{sign\_conditions}(\mathit{basis})$
    \State $R \gets eliminate\_redundancy(S)$
    \For{each constraint $i$ in $R$}
      \State $D_{\mathrm{next}} \gets \mathit{compute\_next(R,i)}$
      \State $W \gets W \cup \{ (R, D_{\mathrm{next}}) \}$
    \EndFor
    \State $\mathit{regions} \gets \mathit{regions} \cup \{ (R, X^*) \}$
    \State $R_{\mathrm{cov}} \gets R$
  \EndIf
  \If{$\neg \mathit{are\_adjacent}(R_{\mathrm{from}},R_{\mathrm{cov}})$}
    \State $D' \gets \mathit{midpoint}(R_{\mathrm{from}},R_{\mathrm{cov}},D)$
    \State $W \gets W \cup \{ (R_{\mathrm{from}}, D') \}$
  \EndIf
\EndWhile
\State \textbf{return}($\mathit{regions}$)
\EndProcedure
\Statex
\Procedure{$\mathit{is\_covered}$}{$D,\mathit{regions})$}
\For{$(R,X^*) \in \mathit{regions}$}
   \If{$D$ covered by $R$}
     \State \textbf{return}($R$)
   \EndIf
\EndFor
\State \textbf{return}(\none)
\EndProcedure
\end{algorithmic}
% \end{multicols}

\end{algorithm}

%%% Local Variables:
%%% ispell-local-dictionary: "en"
%%% TeX-master: "main"
%%% End:

\section{Parallel algorithms}
\label{sec:para}

\subsection{Parallel redundancy elimination}
A polyhedron may be specified by a redundant system of inequalities, meaning that some inequalities can be discarded without changing the polyhedron.
The first step is to eliminate syntactic redundancies --- constraints simplified into {\true} or {\false}, or subsumed by another
(\eg, $2x+2y \leq 2$ subsumes $x+y \leq 2$); a constraint of the form $0 \leq -1$ after simplification makes the polyhedron empty; a constraint of the form $0 \leq 1$ after simplification is to be discarded;
if two constraints are of the form $C^T X \leq B_1$ and  $C^T X \leq B_2$ where $B_1 \leq B_2$, then the latter is to be discarded (note that this involves putting the vector of rationals $C$ in canonical form: flushing denominators and removing common factors, so that \eg, $2x + 2y \leq 3$ can be discarded as subsumed by $x + y \leq 1$).

The general case of redundancy elimination is harder. Checking that an inequality $C$ is redundant with respect to other inequalities $C_1 \land \dots \land C_n$ boils down to finding a vector $D$ such that $D$ satisfies $C_1 \land \dots \land C_n$ but not $C$: such a vector exists if and only if the inequality is irredundant.
This is a pure satisfiability problem in linear programming (it uses a strict inequality $\neg C$, but this can be dealt with).
Note that if an inequality is irredundant, a vector $D$ not satisfying that inequality is provided: this is handy since $\mathit{eliminate\_redundancy}(S)$ is to, in addition to removing constraints, provide a set of vectors each violating one constraint but not the others.

\begin{algorithm}[h]
  \caption{Parallel redundancy elimination.}
  \label{algo:redundancy}
  % \begin{multicols}{2}
  \begin{algorithmic}
    \State $\mathit{is\_redundant} \gets \mathit{make\_array}(n,\false)$
    \For{$i \in 0\dots n-1$}
      \State $F \gets \{ \neg C_i \}$
      \For{$j \in 0\dots n-1$}
        \State \textbf{atomic} $r \gets \mathit{is\_redundant}[j]$
        \If{$j \neq i \land r$}
          \State $F \gets F \cup \{ C_j \}$
        \EndIf
      \EndFor
      \If{$\neg \mathit{check\_sat}(F)$}
         \State \textbf{atomic} $\mathit{is\_redundant}[i] \gets \true$
      \EndIf
    \EndFor
  \end{algorithmic}
  % \end{multicols}
\end{algorithm}

A sequential algorithm for removing redundant constraints thus considers each constraint in sequence, and tests its irredundancy with respect to all the other remaining constraints that have not been shown to be redundant yet.
This can also be done in parallel, as in Algorithm~\ref{algo:redundancy}.
The only requirement is that the table marking which constraints have already been found to be redundant should support atomic accesses.
We applied the same parallelization scheme to our more refined ``ray-tracing'' redundancy elimination algorithm~\cite{DBLP:conf/vmcai/MarechalP17}.

We can use parallel redundancy elimination within parallel parametric linear programming.
It helps in some cases where the coarse-grained parallelism of the parametric linear programming solver is limited due to the lack of tasks that can be executed in parallel, \eg, projections of polyhedra in smaller dimensions and few faces in the result;
otherwise it does not help and hampers tuning.

  \subsection{Parallel parametric linear programming}
We implemented two variants of task-based coarse-grained parallelism: one using Intel's Thread Building Blocks (TBB),%
\footnote{\url{https://www.threadingbuildingblocks.org/}}
the other using OpenMP tasks~\cite{OpenMP_4.5}.

Algorithm~\ref{algo:PLP} boils down to executing tasks taken from a working set, which can themselves spawn new tasks.
In addition to the working set, it uses a shared data structure: the set $\mathit{regions}$ of regions already seen, used:
\begin{inparaenum}[i)]
\item for checking whether a vector $D$ belongs to a region already covered ($\mathit{is\_covered}$);
\item for checking adjacency of regions;
\item for appending new regions found.
\end{inparaenum}

We implemented it as a concurrent extensible array, with a ``push at end'' operation, random read accesses, and iteration:
either \verb|tbb::concurrent_vector|,
or our simple lock-free implementation based on an array with a large, statically defined maximal capacity, using atomic operations for increasing the current number $n_{\mathrm{fill}}$ of items;
the latter has the advantage of not needing TBB.
There is a little subtlety involved here, in both implementations:
$n_{\mathrm{fill}}$ is incremented atomically before the item is pushed, so that if two items are pushed concurrently, they are pushed to different slots.
However, not all items in slots $0 \dots n_{\mathrm{fill}}-1$ are ready for reading: there may be items currently being written in the last slots.
We therefore need a second index, $n_{\mathrm{ready}} \leq n_{\mathrm{fill}}$, such that all slots $0 \dots n_{\mathrm{ready}}-1$ contain data ready for reading,
which is updated as in Algorithm~\ref{algo:push_region}.

\begin{algorithm}[h]
  \caption{Concurrent push on the shared region structure.}
  \label{algo:push_region}
  % \begin{multicols}{2}
  \begin{algorithmic}
    \Procedure{push\_region}{$R$}
  \State \textbf{atomic} ($i \gets n_{\mathrm{fill}}$; $n_{\mathrm{fill}} \gets n_{\mathrm{fill}} +1$)
  \State $\mathit{regions}[i] \gets R$
  \While{$n_{\mathrm{ready}} < i$}
    \State possibly use a condition variable instead of spinning
  \EndWhile \Comment{$n_{\mathrm{ready}} = i$}
  \State \textbf{atomic} $n_{\mathrm{ready}} \gets i+1$
  \EndProcedure
  \end{algorithmic}
  % \end{multicols}
\end{algorithm}

The working set is managed differently depending on whether we use TBB or OpenMP.
For TBB, we use \verb|tbb::parallel_do|, which dynamically schedules tasks from the working set on a number of threads and allows dynamically adding tasks to the working set.
For OpenMP, when we run a task, we collect all the tasks $(R,D)$ that it generates for spawning, and we spawn them at the end.

The resulting implementation however had disappointing performance.
In fact, we obtained better performance using a naive OpenMP implementation that collected all tasks to spawn, spawned them and waited for them to complete, using a barrier, before the next round of spawning!

We identified the reason. It was frequent that the working set contained two tasks $(R_1,D_1)$ and $(R_2,D_2)$ such that the regions generated from $D_1$ and $D_2$ were the same.
In that case, there were two tasks that each solved a linear program, found the same basis, reconstructed exactly the solution and the parametric objective function in that basis, etc.
The workaround was to add a hash table (either a \verb|tbb::concurrent_unordered_set| or a normal hash table, protected by a mutex) that stores the set of bases (each basis being identified by the ordered set of its nonbasic variables) that have been processed or are currently under processing.
A task aborts after solving the floating-point linear program if it finds a basis identical to one already in the table.

\begin{algorithm}[h]
  \caption{Task for parallel linear programming solver.\label{algo:pplp}}

  $\mathit{push\_tasks}$ adds new tasks to those to be processed;
  its implementation is different under TBB and OpenMP.

  $\mathit{test\_and\_insert}(T,x)$ checks whether $x$ already belongs to the hash table $T$, in which case it returns $\true$; otherwise it adds it and returns $\false$. This operation is atomic.
  % \begin{multicols}{2}
  \begin{algorithmic}
  \Procedure{process\_task}{$(R_{\mathrm{from}},D)$}
  \State $R_{\mathrm{cov}} \gets \mathit{is\_covered}(D,\mathit{regions})$
  \If{$R_{\mathrm{cov}} == \none$}
    \State $\mathit{basis} \gets \mathit{float\_lp}(A,B,C(D))$
    \If{$\neg \mathit{test\_and\_insert}(\mathit{bases}, \mathit{basis})$}
      \State $X^* \gets \mathit{exact\_point}(\mathit{basis})$
      \State $o \gets \mathit{exact\_objective}(\mathit{basis})$
      \If{$\neg (X^* \geq 0 \land o \leq 0)$}
        \State $(\mathit{basis},X^*) \gets \mathit{exact\_lp}(A,B,C(D))$
      \EndIf
      \State $S \gets \mathit{sign\_conditions}(\mathit{basis})$
      \State $R \gets eliminate\_redundancy(S)$
      \For{each constraint $i$ in $R$}
        \State $D_{\mathrm{next}} \gets \mathit{compute\_next(R,i)}$
        \State $\mathit{push\_tasks}(D_{\mathrm{next}})$
      \EndFor
      \State $\mathit{push\_region(R, X^*)}$
      \State $R_{\mathrm{cov}} \gets R$
    \EndIf
  \EndIf
  \If{$\neg \mathit{are\_adjacent}(R_{\mathrm{from}},R_{\mathrm{cov}})$}
    \State $D' \gets \mathit{midpoint}(R_{\mathrm{from}},R_{\mathrm{cov}},D)$
    \State $W \gets W \cup \{ (R_{\mathrm{from}}, D') \}$
  \EndIf
  \EndProcedure
\Statex
\Procedure{$\mathit{is\_covered}$}{$D,\mathit{regions})$}
\For{$i \in 0\dots n_{\mathrm{ready}}-1$}
   \Comment{$n_{\mathrm{ready}}$ to be read at every loop iteration}
   \State $(R,X^*) \gets \mathit{regions}[i]$
   \If{$D$ covered by $R$}
     \State \textbf{return}($R$)
   \EndIf
\EndFor
\State \textbf{return}(\none)
\EndProcedure
  \end{algorithmic}
  % \end{multicols}
\end{algorithm}

The overall algorithm is simple: create an initial task $(\none,D_0)$ and then run the tasks over available threads as long as there are uncompleted tasks (Algorithm~\ref{algo:pplp}).

The number of tasks running to completion (not aborted early due to a test) is the same as the number of generated regions.
Thus, if, geometrically, the problem does not have many enough regions in comparison to the number of available threads of execution, its parallelism is intrinsically limited.

The $\mathit{is\_covered}(D,\mathit{regions})$ loop can be easily parallelized as well.
We opted against it as it would introduce a difficult-to-tune second level of parallelism.

\begin{figure}[h!!]
  \subfloat[Task graph with 1 thread\label{fig:P_29_15_16_3_0_nproj7_nthreads1}]{
    \includegraphics[width=\textwidth]{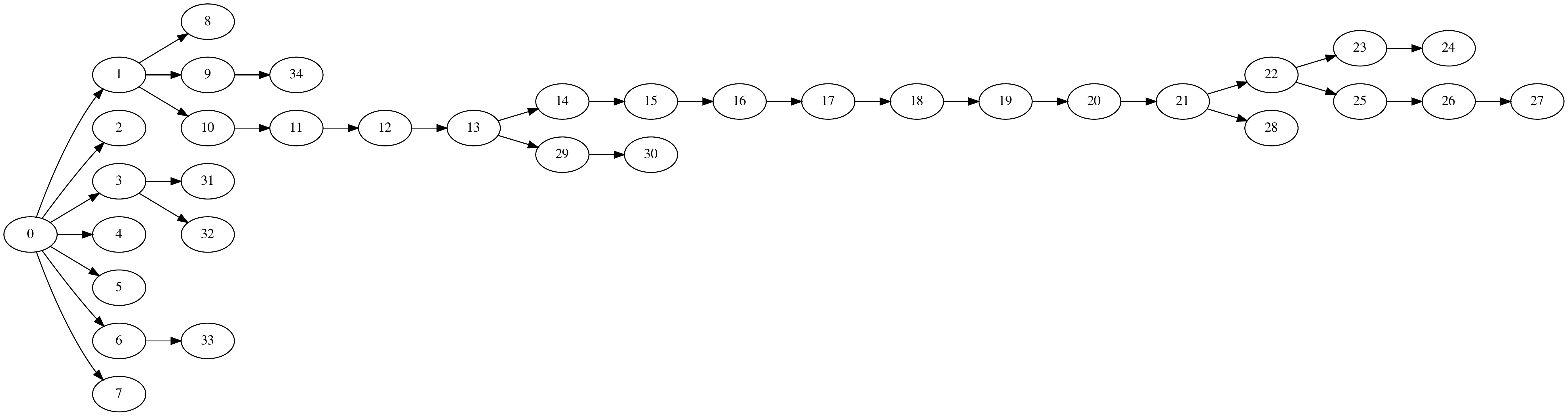}
  }\\
\begin{center}
  \subfloat[Task graph with multiple threads\label{fig:P_29_15_16_3_0_nproj7_nthreads30}]{
    \includegraphics[width=.59\textwidth]{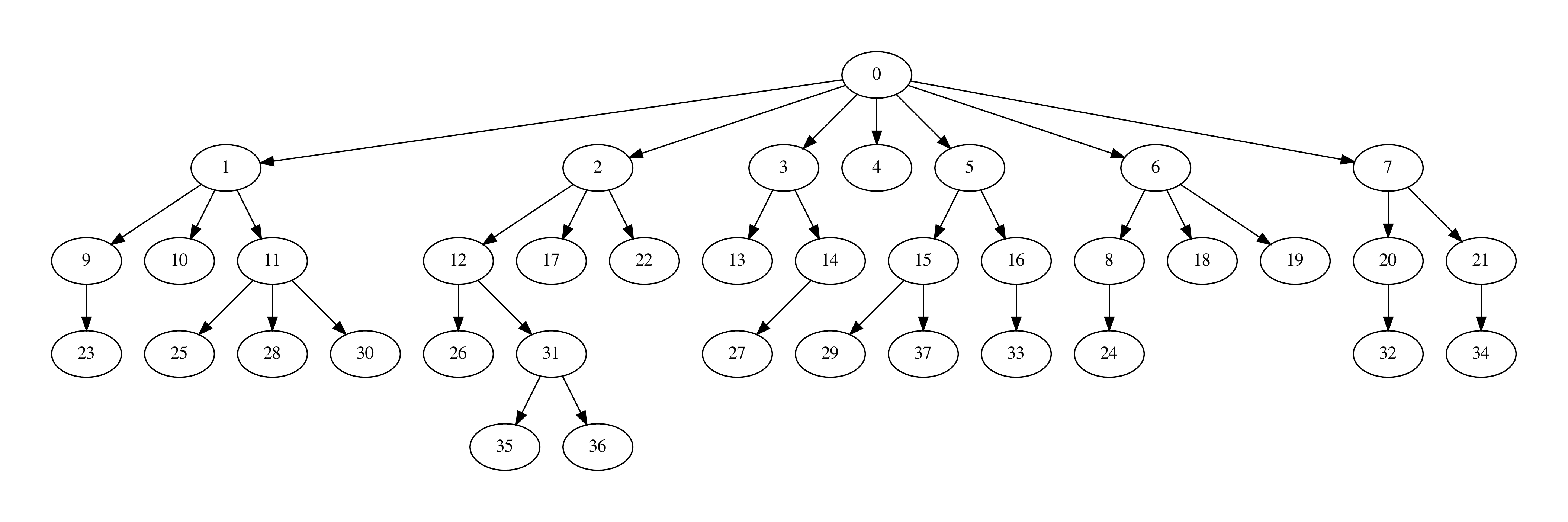}
  }
\end{center}
  \caption{Generation graph of the regions from one typical polyhedron, computed with 1 thread and 30 threads.
    The region graphs, depending on overlaps etc., are different; the numbers in both trees have no relationship.}
  \label{fig:small_polyhedra_spanning_trees}
\end{figure}

Figure \ref{fig:small_polyhedra_spanning_trees} presents the tasks
generated by the sequential and the parallel algorithms on a real
example, and their dependencies. Figure
\ref{fig:P_29_15_16_3_0_nproj7_nthreads1} show how the sequential
algorithm handle tasks: as long as the tasks generate subtasks, these
subtasks are computed. For instance, task 10 has a long series of
descendants that are computed sequentially. Figure
\ref{fig:P_29_15_16_3_0_nproj7_nthreads30} show how the tasks are
handled in parallel: after an initial task 0 that generates a set of
subtasks (tasks 1 to 9), these subtasks are computed in parallel, as
well as the subtasks they generate.

This figure also illustrates the parallelism extracted from the
computation, on a polyhedron involving 29 constraints and 16
variables. In particular, we can see on
Fig.~\ref{fig:P_29_15_16_3_0_nproj7_nthreads30} the number of 
parallel tasks and their dependencies. For instance, task 5 is
generated by task 0 and generates tasks 15 and 16. We can see that, at the
beginning of the computation, task 0 generates 7 tasks (i.e., regions
to compute). Hence, 7 cores will be used by the parallel
algorithm. Since the computation time of each task varies a lot
between tasks, the second level of the tree is not necessarily
executed at the same time.

%%% Local Variables:
%%% ispell-local-dictionary: "en"
%%% TeX-master: "main"
%%% End:

\section{Performance evaluation}
\label{sec:xp}

\begin{figure}[ht]
  \begin{center}
    \subfloat[2 dimensions projected]{
      \includegraphics[width=0.48\textwidth]{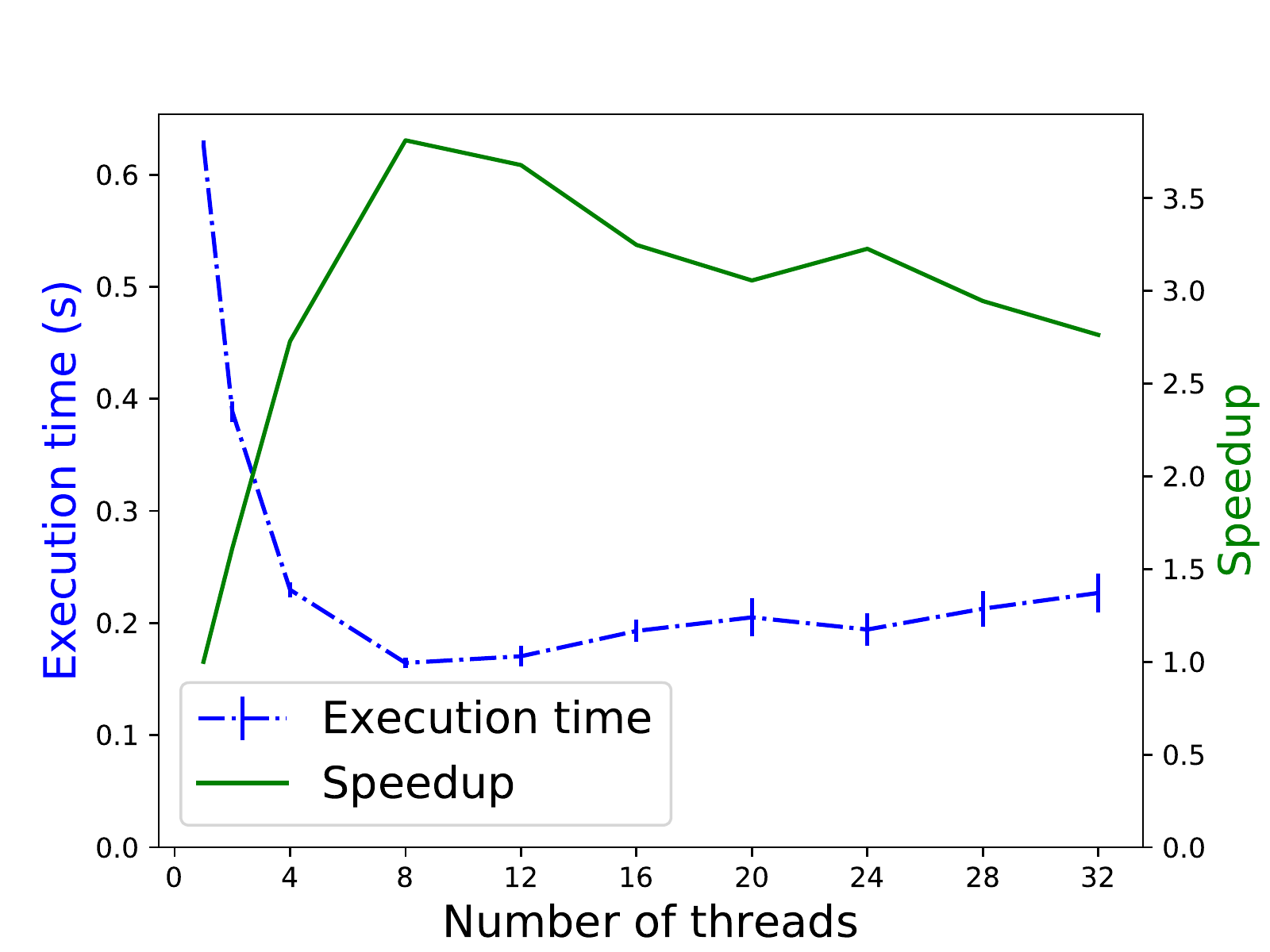}
      \label{fig:xp:9_0_20_16_2}
    }\hfill
    \subfloat[5 dimensions projected]{
      \includegraphics[width=0.48\textwidth]{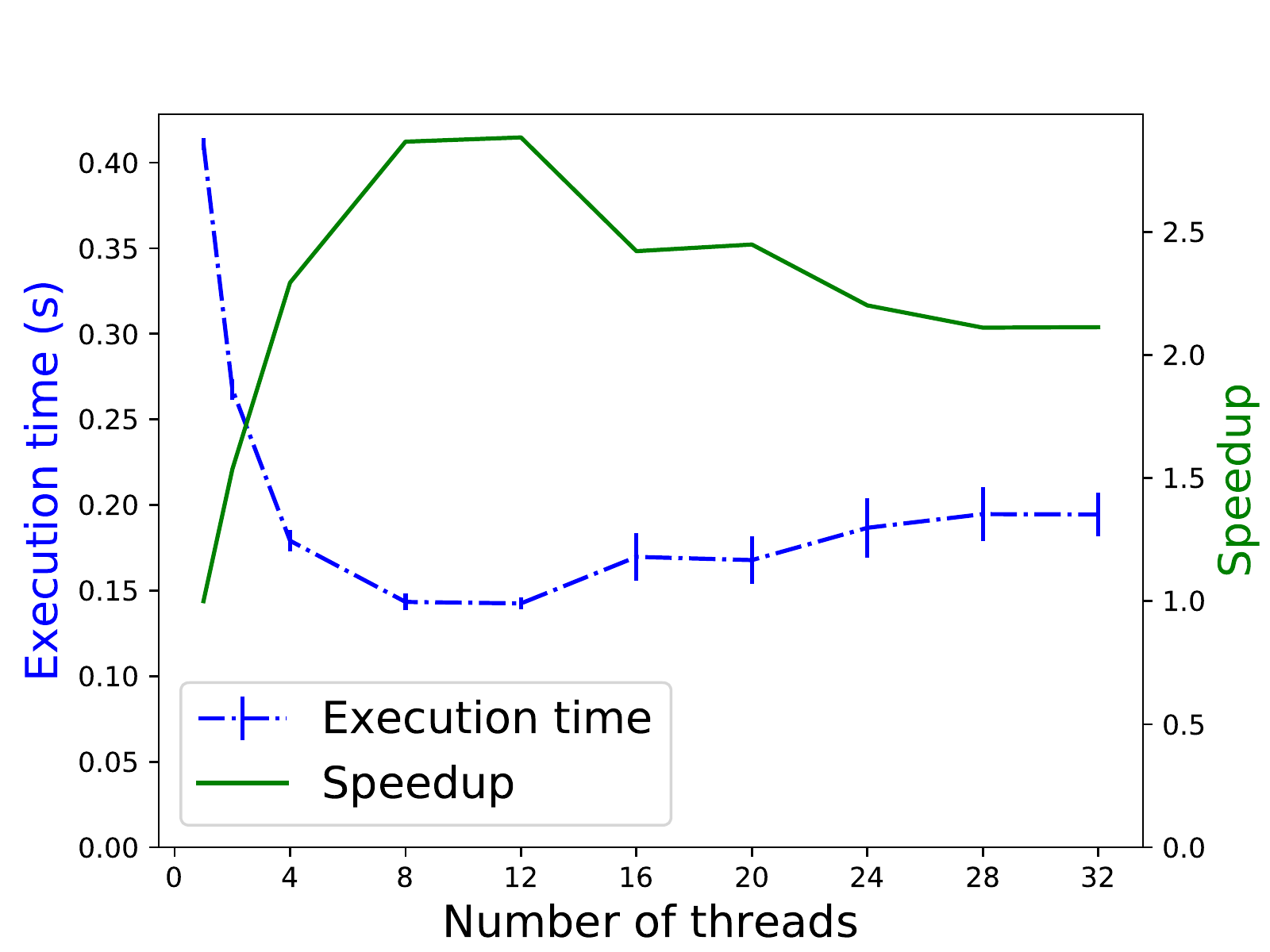}
      \label{fig:xp:9_0_20_16_5}
    }
    \caption{Computation time and speedup for 20 projections of polyhedra with 9 constraints with no redundant constraints and 16 variables, on Paranoia (20 hyperthreaded cores, OpenMP). Each parametric linear program has 2--36 regions}
    \label{fig:xp:9_0_20_16}
  \end{center}
\end{figure}

We implemented our parallel algorithms in C++, with three alternate schemes selectable at compile-time: no parallelism, OpenMP parallelism, TBB.
We use:
\begin{description}
\item[Eigen] for floating-point matrix computations outside of linear programming.
  Eigen supports internal OpenMP parallelism; we disabled it since these computations take very little time in our overall execution and using it would have entailed tuning for two levels of parallelism. We used Eigen 3.3.4.

\item[GLPK] for floating-point linear programming.
  This library was not thread-safe, meaning it was impossible to solve linear programs in multiple threads at the same time.
  We suggested to the maintainer making one global variable thread-local, which solved the problem and was made the default as of GLPK version 4.61; we used 4.63.
  GLPK has no internal parallelism.

\item[Flint] for computations on rationals and rational matrices.
  This library is designed to be thread-safe, but does not use parallelism by itself, at least for the operations that we use. We used Flint 2.5.2.
\end{description}

All the benchmarks were run on the Paranoia cluster of Grid'5000
\cite{grid5000} and on a server called Pressembois. Paranoia features
8 nodes, each equipped with 2 Intel Xeon
E5-2660v2 CPUs (10 cores or 20 threads/CPU, 40 threads per node) and 128 GB of RAM. Although the network
was not used for these experiments, each node as two 10 Gbps and one 1
Gbps Ethernet NICs. The code was compiled using GCC 6.3.1 and OpenMP
4.5 (201511). The nodes run a Linux Debian Stretch environment with a
4.9.0 kernel. Pressembois fetures 2 Intel(R) Xeon(R) Gold 6138 CPU
(20 cores or 40 threads/CPU, 80 threads per node) and 192 GB of RAM. It runs a 4.9 Linux
kernel, and we used GCC 6.3 to compile the code. Every experiment was
run 10 times, and the plots presented in this section provide the
average and standard deviation. Paranoia was used for the OpenMP
experiments, whereas Pressembois was used for the TBB experiments.

We evaluated our parallel parametric linear programming implementation by using it to project polyhedra.
This is a very fundamental operation on polyhedra, since it is used for computing forward images (image of a polyhedron by an affine linear transformation), and may also be used for computing convex hulls, although there exists a more direct approach for that.

We used a set of typical polyhedra, with different characteristics:
numbers of dimensions and of constraints, sparsity,
number of dimensions to be projected.
Here we present a subset of these benchmarks.
Each benchmark comprises 50 to 100 polyhedra.

Our polyhedra were randomly generated. The reason is that it is difficult to obtain polyhedra typically used for the target application (static analysis): because libraries based on the dual description approach behave exponentially with respect to the dimension, static analysers are typically designed to keep low the dimension of the polyhedra, at the expense of analysis precision.%
\footnote{There is a chicken-and-egg problem there: static analysis tools do not use general convex polyhedra or only in low dimension due to the cost of the dual description in higher dimension, thus there is little incentive to develop libraries more efficient in higher dimension. Designers of such libraries then lack higher dimension examples from static analysis.}

On problems that have only few regions, not enough parallelism
can be extracted to exploit all the cores of the machine. For
instance, Figure \ref{fig:xp:9_0_20_16} presents two experiments on 2
to 36 regions using the OpenMP version. It gives an acceptable speedup on a few cores (up to
10), then the computation does not generate enough tasks to keep the
additional cores busy.

\begin{figure}[h]
  \begin{center}
    \subfloat[2 dimensions projected, 4 redundant constraints]{
      \includegraphics[width=0.48\textwidth]{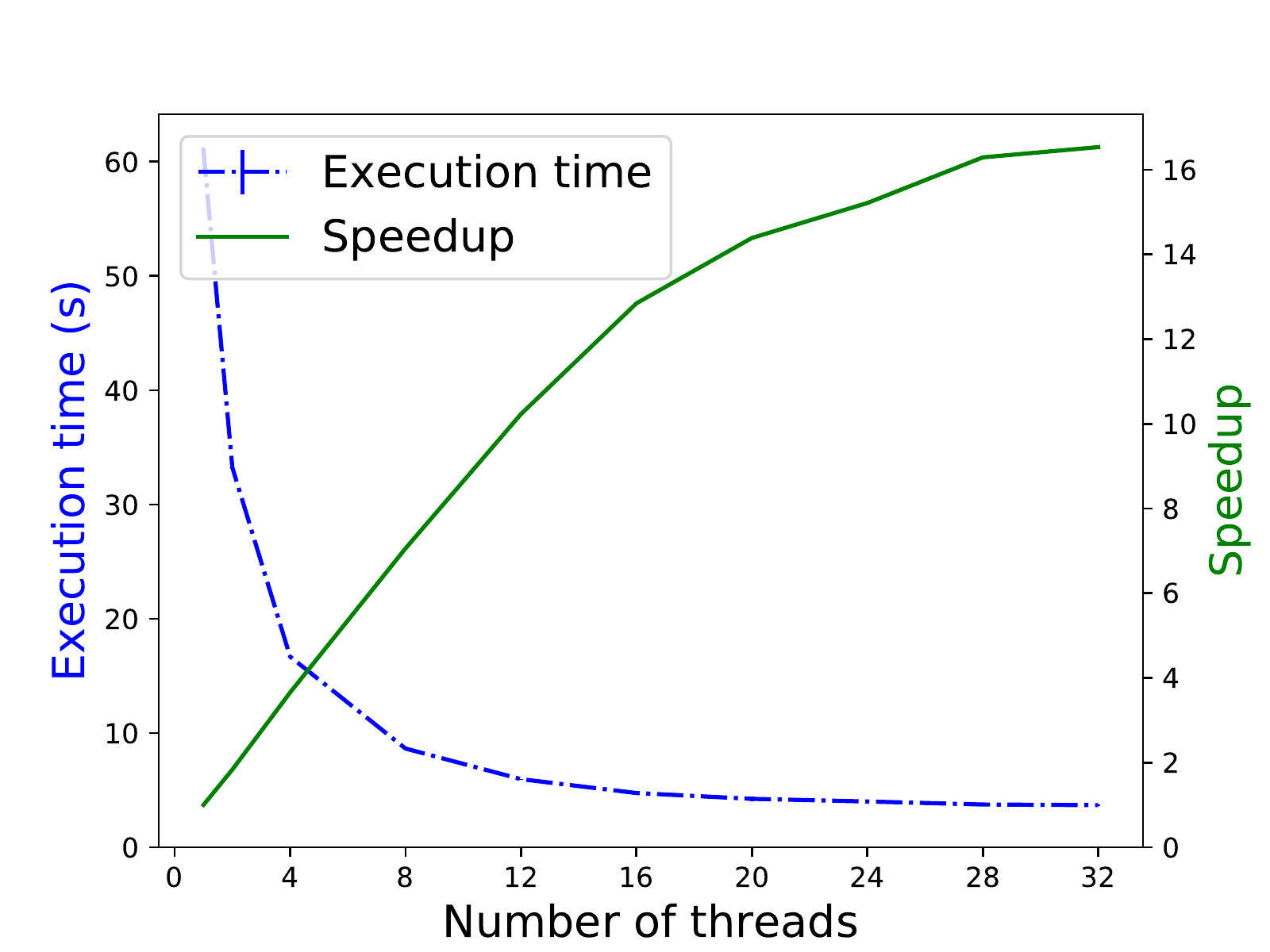}
      \label{fig:xp:24_4_10_6_2}
    }\hfill
    \subfloat[5 dimensions projected, 4 redundant constraints]{
      \includegraphics[width=0.48\textwidth]{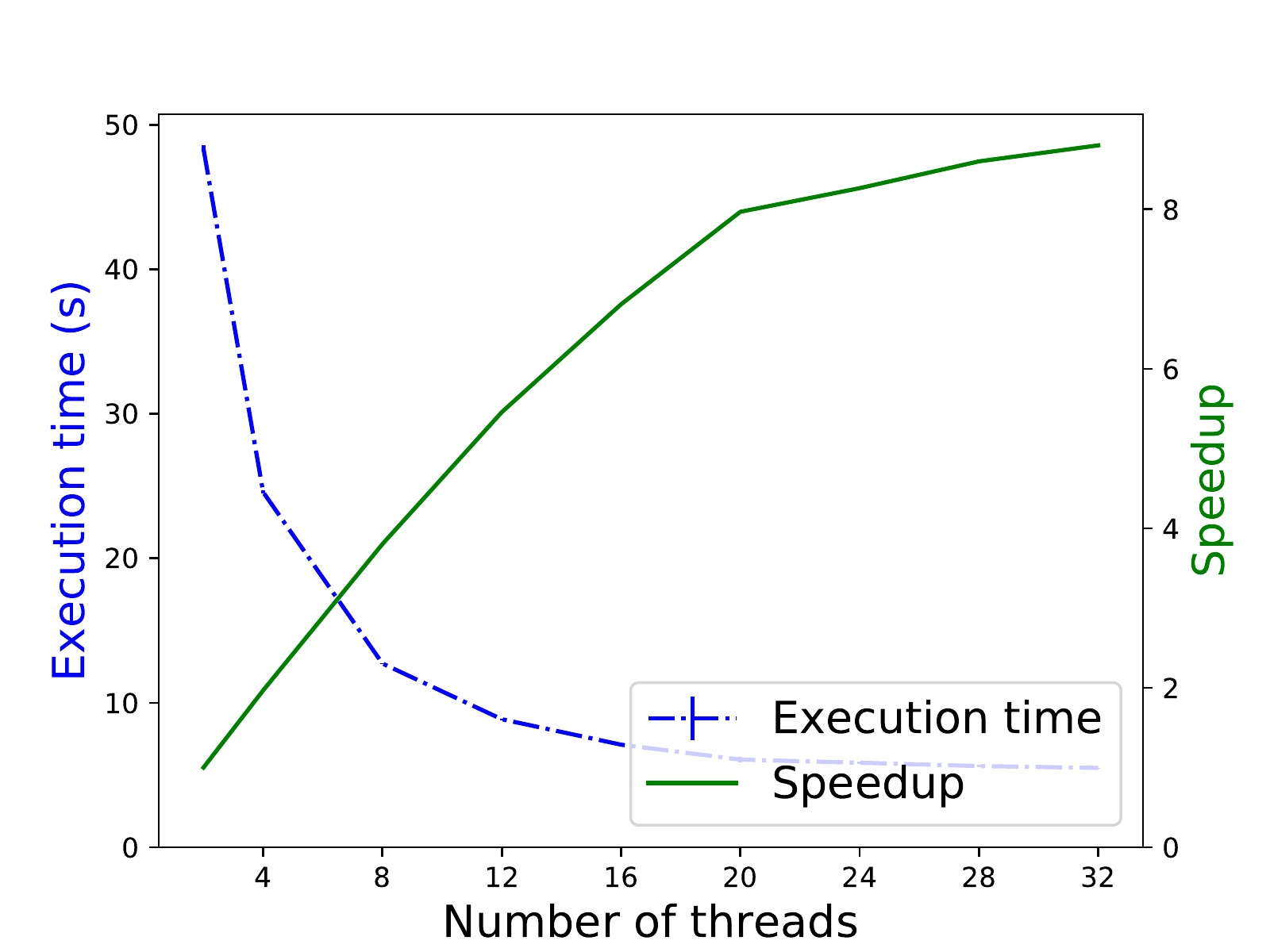}
      \label{fig:xp:24_4_10_6_5}
    }

    \subfloat[2 dimensions projected, 12 redundant constraints]{
      \includegraphics[width=0.48\textwidth]{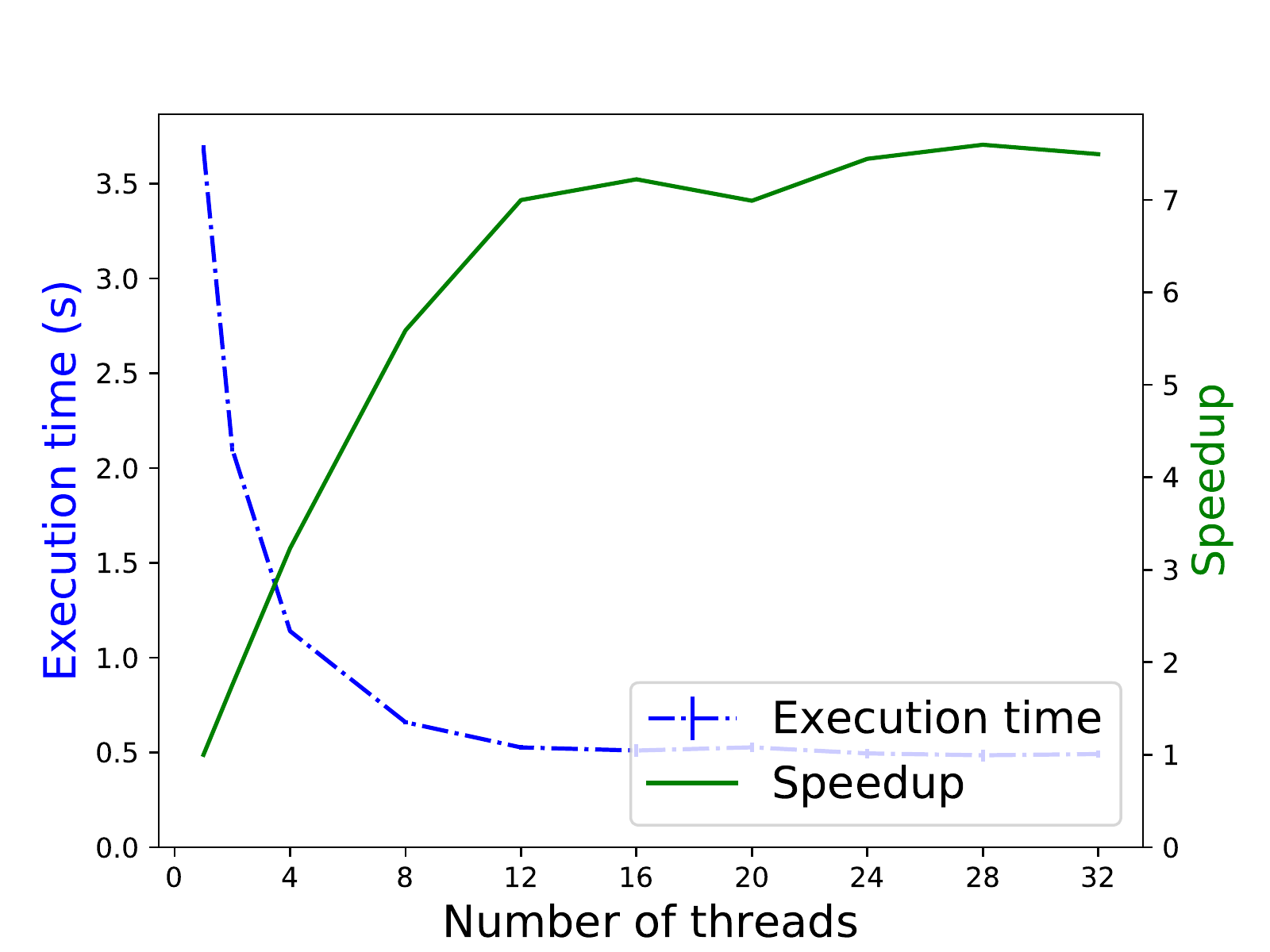}
      \label{fig:xp:24_12_10_6_2}
    }\hfill
    \subfloat[5 dimensions projected, 12 redundant constraints]{
      \includegraphics[width=0.48\textwidth]{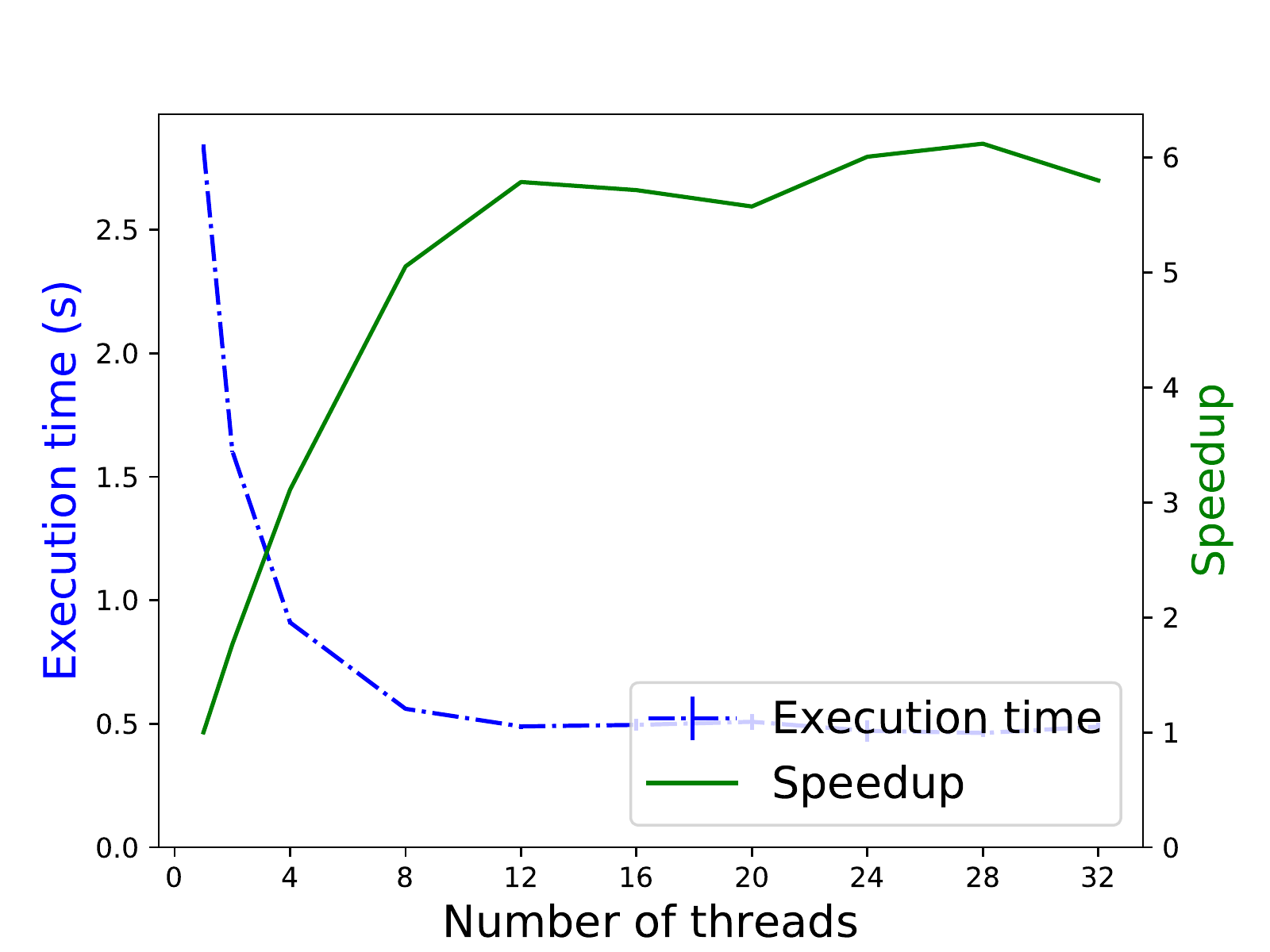}
      \label{fig:xp:24_12_10_6_5}
    }
    \caption{Computation time and speedup for different numbers of projections of polyhedra with 10 variables and variable numbers of redundant constraints, on Paranoia (20 hyperthreaded cores, OpenMP). Each parametric linear program has 8--764 regions.}
    \label{fig:xp:24_4_10_6}
  \end{center}
\end{figure}

As expected, when the solution has a large number of
regions, computation scales better. Figure \ref{fig:xp:24_4_10_6}
presents the performance obtained on polyhedra made of 24 constraints,
%(4 and 12 redundant regions),
involving 8 to 764 regions and using the OpenMP version. The speedup
is sublinear. This is likely due to the synchronizations when
tasks are created (lookup in the hash table), some contention on
shared data structures and task management.

\begin{figure}[t]
  \subfloat[OpenMP on Paranoia]{
    \includegraphics[width=.49\textwidth]{times_omp_P_120_0_50_0_proj1_rennes_be0be063}
    \label{fig:xp:120omp}
  }
  \subfloat[TBB on Pressembois]{
    \includegraphics[width=.49\textwidth]{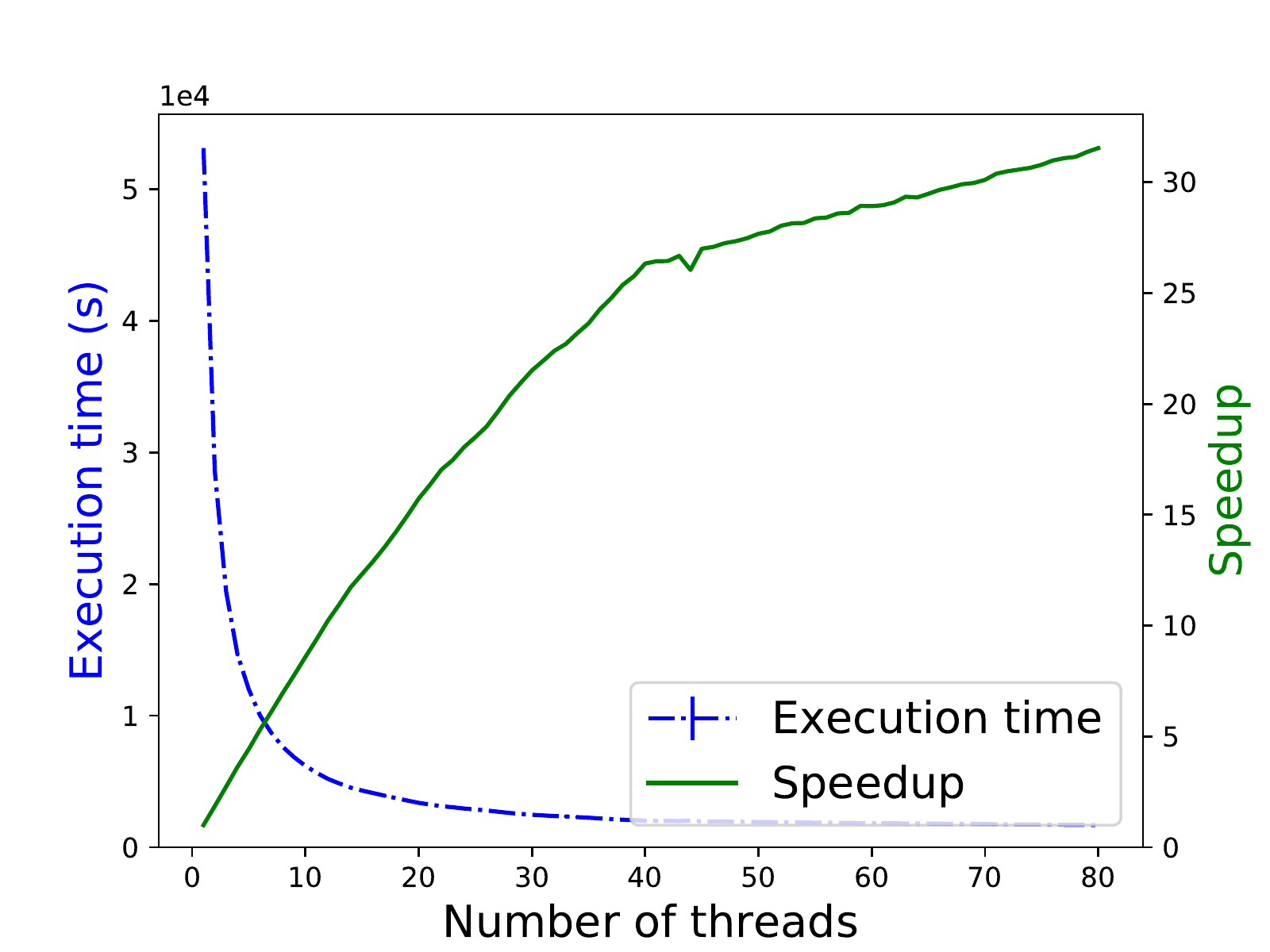}
    \label{fig:xp:120tbb}
  }
  \caption{120 constraints, 50 variables, 1 dimension projected, 3459--3718 regions.}
  \label{fig:xp:120}
\end{figure}

On larger polyhedra, with 120 constraints and 50 variables, we can
see however that the speedup is close to a linear one with OpenMP
as well as with TBB (Figure \ref{fig:xp:120}).

\begin{figure}[ht]
\begin{minipage}{.48\textwidth}
{\centering\includegraphics[width=\textwidth]{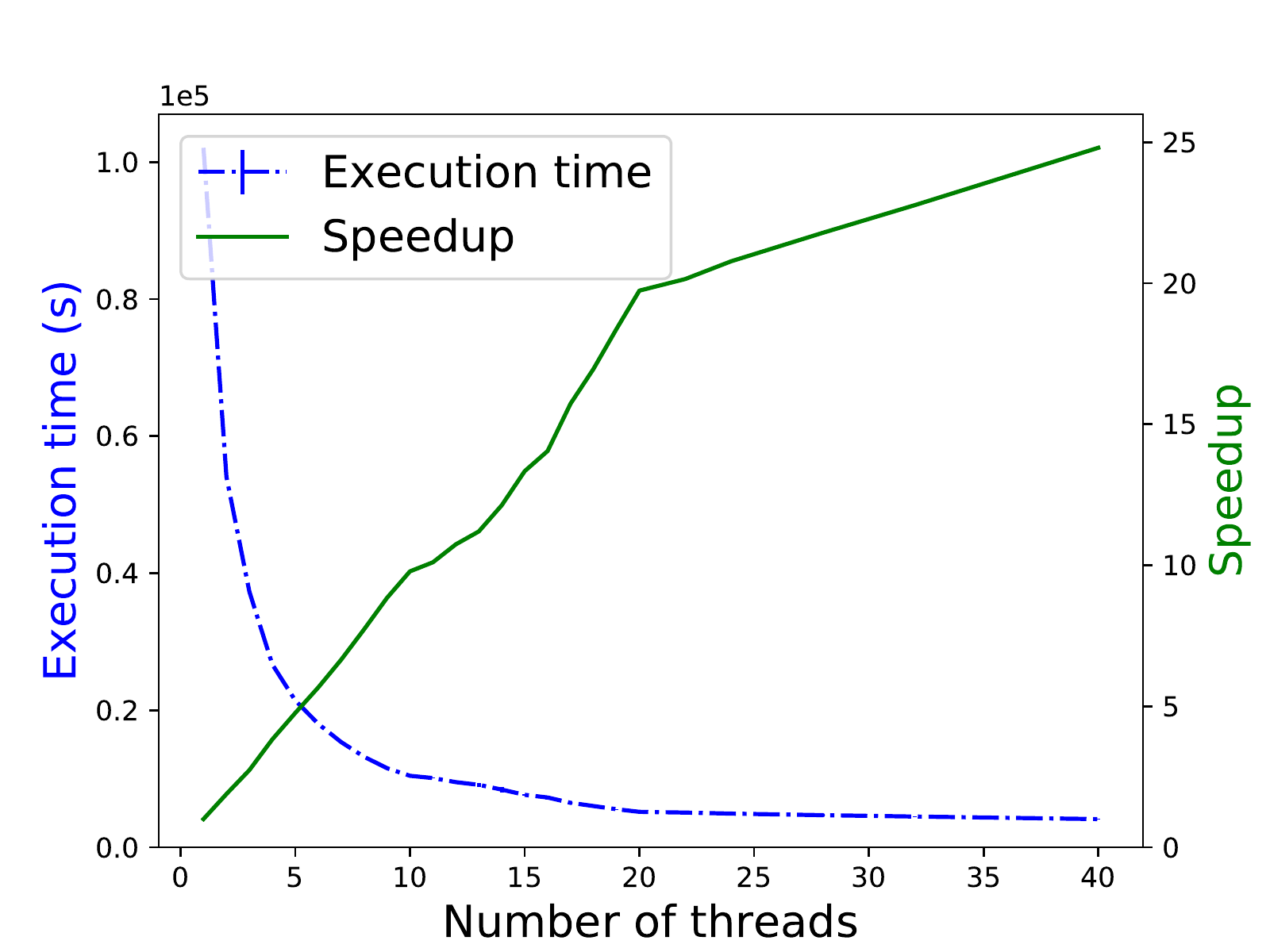}}
\caption{Computation time and speedup for 50 projections of polyhedra in dimension 120, on Paranoia (20 hyperthreaded cores, OpenMP). Each parametric linear program has 3460--3715 regions.}
\label{fig:large_polyhedra}
\end{minipage}%
\hfill %
\begin{minipage}{.48\textwidth}
  {\centering\includegraphics[width=\textwidth]{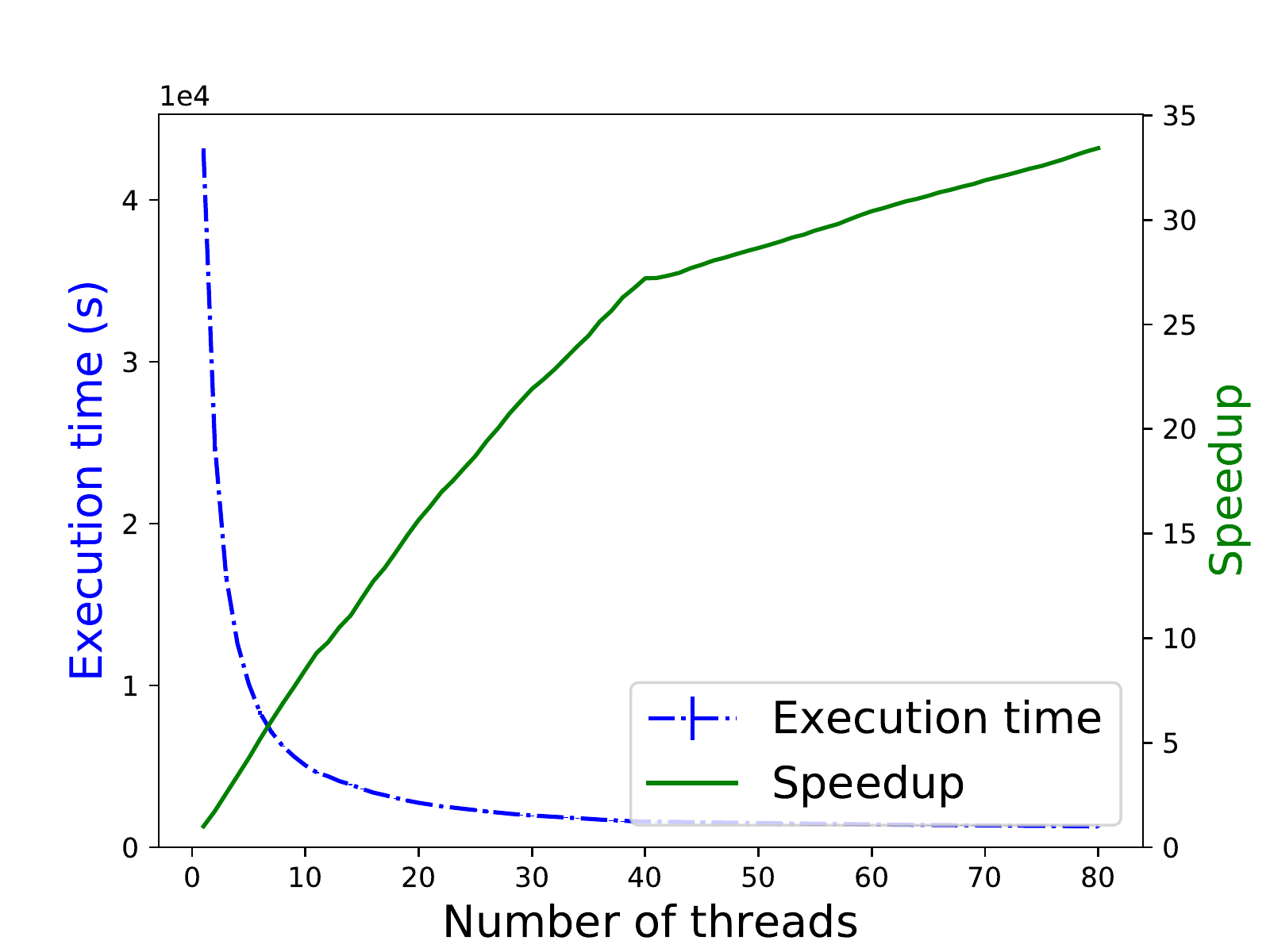}}
\caption{Computation time and speedup for 50 projections of polyhedra in dimension 100, on Pressembois (40 hyperthreaded cores, TBB).}
\end{minipage}
\end{figure}

\begin{figure}[ht]
\begin{center}
  \begin{minipage}{.48\textwidth}
{\includegraphics[width=\textwidth]{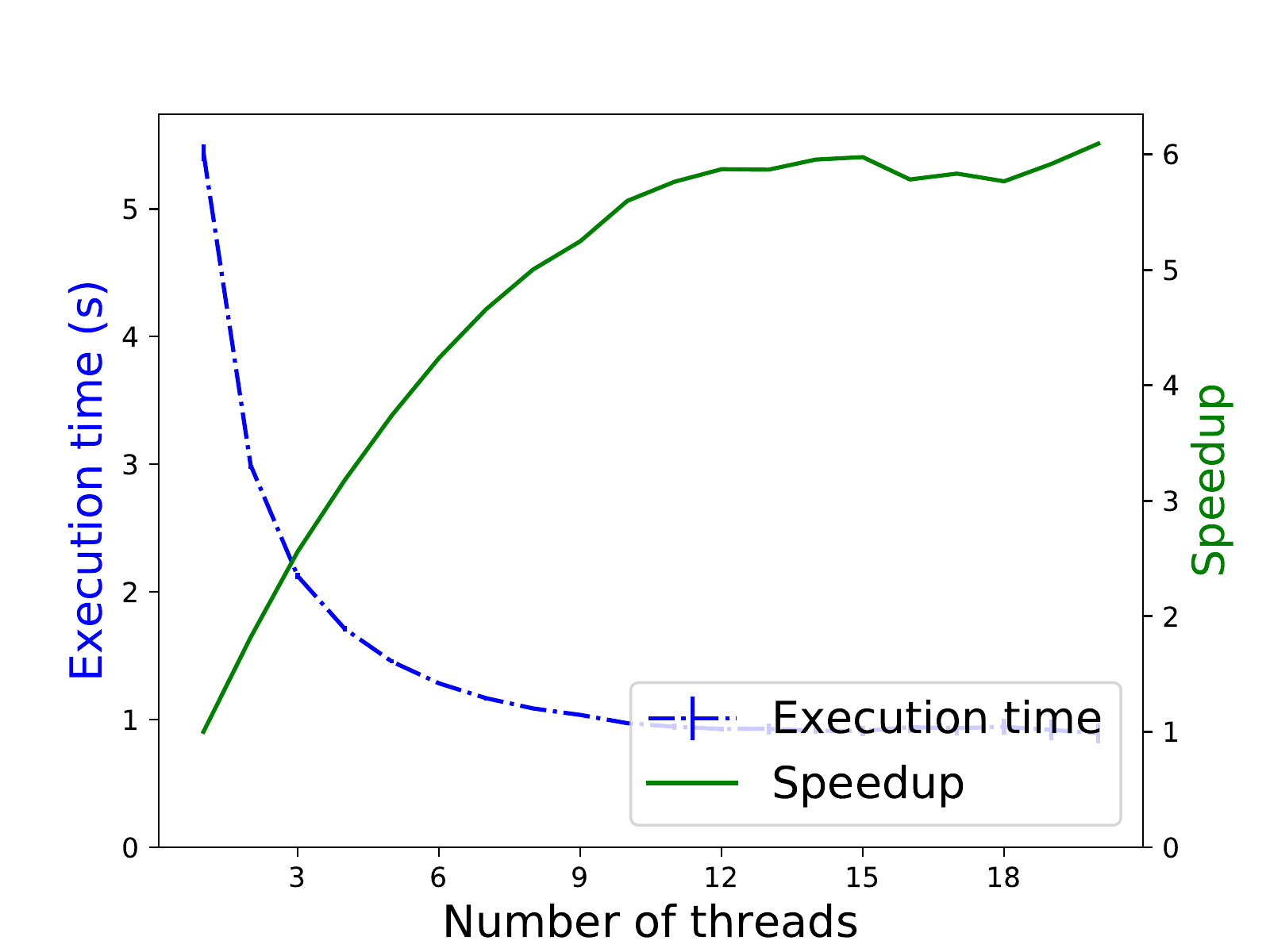}}
\caption{Computation time and speedup for smaller, simpler polyhedra: 16 projections in dimension 29, on Paranoia (20 cores, OpenMP).}
\label{fig:small_polyhedra}
\end{minipage}
\end{center}
\end{figure}

The speedup is good for larger problems with many regions (which can
be computed independently from each other):
the number of tasks is larger than the number of available cores, hence allowing
an efficient parallel computation (Fig.~\ref{fig:large_polyhedra}).
When the problem does not have enough regions to allow the algorithm
to extract enough parallelism to use all the available cores, the
speedup is bounded by how many independent tasks are generated. We
have seen an example presented Figure
\ref{fig:small_polyhedra_spanning_trees}, where at the
beginning of the computation, 7 cores at most can be used in
parallel. Hence, a plateau occurs when the problems do not have enough regions (Fig.~\ref{fig:small_polyhedra});
the limited width of the region graph then limits parallelism.
%(Fig.~\ref{fig:small_polyhedra_spanning_trees}).

%\begin{figure}[t]
%  \subfloat[Task graph with 1 thread\label{fig:P_29_15_16_3_0_nproj7_nthreads1}]{
%    \includegraphics[width=\textwidth]{P_29_15_16_3_0_nproj7_nthreads1}
%  }\\
%  \subfloat[Task graph with 30 threads\label{fig:P_29_15_16_3_0_nproj7_nthreads30}]{
%    \includegraphics[width=.59\textwidth]{P_29_15_16_3_0_nproj7_nthreads30}
%  }
%  \subfloat[Performance on Paranoia (20 hyperthreaded cores, OpenMP).]{
%        \includegraphics[width=.39\textwidth]{times_auto_P_29_15_16_3}
%  }
%  \caption{Generation graph of the regions from one typical polyhedron, computed with 1 thread and 30 threads.
%    The region graphs, depending on overlaps etc., are different; the numbers in both trees have no relationship.}
  %  \label{fig:small_polyhedra_spanning_trees}
%\end{figure}

%%% Local Variables:
%%% ispell-local-dictionary: "en"
%%% TeX-master: "main"
%%% End:

\section{Conclusion and future work}
\label{sec:conclu}
We have successfully parallelized computations over convex polyhedra represented as constraints, most importantly parametric linear programming.
Speedups are particularly satisfactory for the most complex cases: polyhedra with many constraints in higher dimension.

The main cause of inefficiency in our current implementation is geometrical degeneracy, which causes overlapping regions.
Several approaches are being studied in this respect, all based on enforcing that, for given parameters, there should be only one optimal basis.%
\footnote{For some of our applications, such as polyhedral projection, there cannot be several optimal vertices for the same parameters, except at region boundaries, so that source of degeneracy is not present. There however remains the other source, that is, several bases for the same vertex.}

The floating-point simplex algorithm is restarted from scratch in each task.
It could be more efficient to store in the task structure the last basis used, or even the full simplex tableau, to start solving from that basis instead of the default initial.
The intuition is that neighboring regions are likely to have similar bases.

We have presented an approach to handle the heterogeneous,
non-predictible computation time, and lack of predictibility of the
domain decomposition. One major challenge of this problem is that each
subtask covers an area of the domain, but the size and shape of this
area is not known before the task is actually computed.
In addition, precautions must be taken to avoid computing the same area multiple
times on multiple tasks. 

Checking whether a vector belongs to a region that has already been
processed is currently implemented following a very simplistic approach: the vector is searched for in every region.
A binary space partitioning region storage could be used instead: build a tree with nodes adorned by hyperplanes, all regions wholly on one side of the hyperplane under the first child, all regions wholly on the other side under the second child, all regions straddling the hyperplane in both branches, and so recursively in the branches.
Then a region covering a vector is searched for by testing, at each node, on which side of the hyperplane the vector lies, and entering the branch.
An appropriate locking scheme would however have to be designed, hence
introducing synchronizations and requiring more collaboration from the
operating system.

Our coarse-grained parallelism for parametric linear programming dependent highly on the geometry of the problem.
If there are too few regions, too few tasks will keep the cores busy and little speedup will be achieved.
One could use two levels of parallelism, with parallel linear programming, parallel exact reconstruction, etc... onn each task, using parallel matrix computations.
One possibility would be to parallelize floating-point linear programming and exact matrix computations,
both handled by external libraries.
These phases are based on matrix operations (either floating-point, rational, or integer modular), so the usual parallelization schemes for matrix computations should apply.
One should keep in mind, however, that our matrices are much smaller than the ones typically used in high performance computing applications.
One option could be to replace GLPK by another library based on BLAS,
such as Coin-OR LP, and use a parallel BLAS implementation such as
GotoBLAS or Intel MKL, using an adaptative, hierarchical parallelism,
as mentionned in \cite{wco17}.
With respect to Flint, we attempted to run certain internal matrix computation loops in parallel, but gave up due to crashes; perhaps this could be done with more careful examination of shared data structures or the collaboration of the library maintainer.

While our approach based on (in)equalities only does not blow up exponentially when the polyhedron is a Cartesian product of simple polyhedra, as opposed to the double description, it would likely benefit from being applied separately to terms of a Cartesian product as opposed to over the full product, perhaps in parallel.
Some computations are performed over dense matrices, and, even if sparseness is exploited, it is easier to exploit it at a coarse level.
We think of combining our approach with a Cartesian decomposition
\cite{DBLP:journals/fmsd/HalbwachsMG06}.

\section*{Acknowledgements}
Experiments presented in this paper were carried out using the Grid'5000 testbed, supported by a scientific interest group hosted by Inria and including CNRS, RENATER and several Universities as well as other organizations (see \url{https://www.grid5000.fr}).

%%% Local Variables:
%%% ispell-local-dictionary: "en"
%%% TeX-master: "main"
%%% End:

\nocite{Motzkin_collected_papers}
\printbibliography
\end{document}